\documentclass[aps,superscriptaddress,english,10pt,noshowpacs,twocolumn]{revtex4}

\usepackage[latin1]{inputenc}
\usepackage{graphicx}
\usepackage{float}
\usepackage{latexsym}
\usepackage{graphicx}
\usepackage{amssymb}
\usepackage{amsmath}
\usepackage{amsfonts}
\usepackage{hyperref}
\usepackage{cool}
\usepackage{dcolumn}
\usepackage{textcomp}
\usepackage{color}
\usepackage{xfrac}

\def\be{\begin{equation}}
\def\ee{\end{equation}}
\newcommand{\bea}{\begin{eqnarray}}
\newcommand{\eea}{\end{eqnarray}}

\usepackage{slashed}

\begin{document}

\title{Strongly-Interacting Ultralight Millicharged Particles}

\author{Stephon Alexander}
\affiliation{Brown Theoretical Physics Center and Department of Physics, Brown University, Providence, RI, USA}
\affiliation{Center for Computational Astrophysics, Flatiron Institute, New York, NY 10003, USA}

\author{Evan McDonough}
\affiliation{Kavli Institute for Cosmological Physics and Enrico Fermi Institute, The University of Chicago, Chicago, IL 60637, USA}
\affiliation{Department of Physics, University of Winnipeg, Winnipeg, MB R3B 2E9, Canada}

\author{David N. Spergel}

\affiliation{Center for Computational Astrophysics, Flatiron Institute, New York, NY 10003, USA}
\affiliation{Department of Astrophysical Sciences, Princeton University, Princeton, NJ 08544, USA}

\begin{abstract}
We consider the implications of an ultra-light fermionic dark matter candidate that carries baryon number. This naturally arises if dark matter has a small charge under standard model baryon number whilst having an asymmetry equal and opposite to that in the visible universe. A prototypical model is a theory of dark baryons of a non-Abelian gauge group, i.e., a dark Quantum Chromo-Dynamics (QCD). For sub-eV dark baryon masses, the inner region of dark matter halos is naturally at `nuclear density', allowing for the formation of exotic states of matter, akin to neutron stars. The Tremaine-Gunn lower bound on the mass of fermionic dark matter, i.e., the dark baryons, is violated by the strong short-range self-interactions, cooling via emission of light dark pions, and the Cooper pairing of dark quarks that occurs at densities that are high relative to the (ultra-low) dark QCD scale.  We develop the astrophysics of these STrongly-interacting Ultra-light Millicharged Particles (STUMPs) utilizing the equation of state of dense quark matter, and find halo cores consistent with observations of dwarf galaxies. These cores are prevented from core-collapse by pressure of the `neutron star', which suggests ultra-light dark QCD as a resolution to core-cusp problem of collisionless cold dark matter. The model is distinguished from ultra-light bosonic dark matter through direct detection and collider signatures, as well as by phenomena associated with superconductivity, such as Andreev reflection and superconducting vortices.
\end{abstract}

\maketitle

\section{Introduction} 

We have but few clues as to the nature of dark matter. The tension between the predictions of the  collisionless cold dark matter paradigm and the observed properties of dwarf galaxies is a potential hint suggesting a maximum characteristic density for dark matter \cite{Burkert, Bullock:2017xww}. The origin of the observed matter anti-matter asymmetry is another potential hint about dark matter properties. Since the CP violation of the standard model is not sufficient to have produced the asymmetry, this stands as one of the few indications of physics beyond the standard model.

 A simple possibility that connects these disparate threads is a dark matter candidate that carries standard model baryon number and self-interacts through dark Quantum Chromo-Dynamics (QCD):
 a Strongly-Interacting Ultralight Millicharged Particle (STUMP).  
This inherits features of both self-interacting dark matter \cite{Spergel:1999mh,Dave:2000ar} and ultra-light bosonic dark matter \cite{Hu:2000ke,Hui:2016ltb,Berezhiani:2015bqa,Ferreira:2018wup,Ferreira:2020fam}, while being easily distinguished from both. This is naturally the case in a baryon-symmetric universe in which the dark matter is `millicharged' under visible baryon number: the cancellation of the visible sector asymmetry demands a high dark matter number density, and hence a small dark matter mass, as more conventionally associated with axions and bosonic dark matter \cite{Marsh:2015xka,Ferreira:2020fam}. 

 Ultralight dark QCD provides an opportunity to probe a region of QCD physics usually relegated to neutron stars, namely, condensate (superfluid, superconducting) phases of QCD. The emergence of these states (along with the strong interactions and dissipation) evades the Tremaine-Gunn bound on ultra-light fermionic dark matter, opening the window for dark baryons of sub-eV masses. Indeed, dwarf galaxies, with central density $\rho \sim 10^{9} M_{\odot}/{\rm kpc}^3\sim (0.13 \, {\rm eV})^4 $ are at nuclear density for $m_{\rm DM} \lesssim 0.13 \, {\rm eV}$, suggesting a central core to galaxies comprised of condensate matter, analogous to a neutron star.

In this {\it Letter} we introduce ultralight dark QCD, wherein dark matter is in the form of dark baryons. Because the dark matter is interactive and can form Cooper pairs, the dark baryons (STUMPs) evade the Tremaine-Gunn bound on ultralight fermionic dark matter \cite{Tremaine:1979we} and limits on collisionless cold fermionic dark matter \cite{Dalcanton:2000hn,Boyarsky:2008ju,Domcke:2014kla,DiPaolo:2017geq,Savchenko:2019qnn,Alvey:2020xsk,Randall:2016bqw}. The model predicts cores to dark matter halos, with a pressure and density profile governed by the equation of state of dense quark matter, analogous to neutron stars. We find cores consistent with observations of dwarf galaxies, suggesting this model as a resolution of the core-cusp problem. We elucidate the key features that distinguish this model from ultralight bosonic dark matter, namely through phenomena of superconductivity, such as superconducting vortices \cite{Eto:2013hoa}, and Andreev reflection \cite{Andreev1964THERMALCO} at the core-halo boundary. In appendices \ref{app:schematic} though \ref{app:early} we provide additional details of the model and supplementary material.

\section{Ultralight Dark QCD}  There is an extensive literature on QCD-like dark matter models \cite{Kribs:2016cew}. These are largely motivated by the observation that the dark matter and visible matter cosmological abundances differ by a factor of $\sim 5$; if one posits that dark matter is a near-copy of visible QCD, this coincidence suggests a cogenesis of visible and dark matter-antimatter asymmetries \cite{Kamada:2012ht,Borah:2019epq,Hall:2010jx,Morrison:2018xla}, and a WIMP-like composite dark matter candidate of mass similar to the proton. In this way, models of baryogenesis can be made predictive for dark matter searches. 

An unexplored regime of dark QCD is a model wherein the dark QCD scale is much lower than in the standard model. Absent a collider measurement of the strong coupling constant from which to anchor the renormalization group flow, there is no prior on the strong coupling scale of an $SU(N)$ gauge theory decoupled from the standard model. In the context of a cogenesis mechanism, the ultralight regime corresponds to the case wherein the dark quarks have visible baryon charge $B_q \ll 1$, such that the requisite number density is high and hence the dark baryon mass is small, $m_B \ll m_{proton}$. This is reminiscent of (electrically) millicharged dark matter \cite{HOLDOM198665,MOHAPATRA1990593}, and concrete models can be built following the recipe proposed in that context: $U(1)_B$ is gauged at a high energy scale $\Lambda_B$, with field strength $B_{\mu \nu}$, the dark quarks are charged under a dark $U(1)$ gauge group with field strength $\tilde{F}_{\mu \nu}$, and the two gauge bosons are kinetically coupled as $\frac{\epsilon}{2} B_{\mu \nu} \tilde{F}^{\mu \nu}$. This endows the dark quarks with a standard model baryon charge of $B_q = \epsilon/3$ (see e.g., \cite{HOLDOM198665,Foot:2010hu}). We build a concrete and anomaly-free model in appendix B.

To understand the physics of this model we must take a tour of the phase diagram of QCD. This is schematically illustrated in Appendix A, Fig.~\ref{fig:QCD-phase-diagram}. 
All phases are governed by the same UV completion, i.e., the QCD Lagrangian,
\begin{equation}
    \mathcal{L} = -\frac{1}{4} {\rm} {\rm Tr}\, G^2+ i \bar{q} \slashed{D} q - m_q \bar{q} q,
\end{equation}
where the trace is over color indices, $q$ is the dark quark,  $m_q$ is the quark mass matrix, and $G$ is the dark gluon field strength tensor. The dark quarks $q$ transform under a baryon number transformation parametrized by $\alpha$ as $q\rightarrow e^{i B_q \alpha}q$ with $B_q \ll 1$, in contrast with the standard model quarks, which transform as $\psi \rightarrow e^{i \alpha/3 }\psi$. The states of matter of the dark QCD theory are distinguished by differing low energy effective actions and spectrum of excitations. 

At low temperatures and density, %the bottom left corner of Fig.~\ref{fig:QCD-phase-diagram},
all excitations are confined within hadrons. This phase is characterized by a condensate $\langle \bar{q} q \rangle \simeq \Lambda_{\rm QCD}^3$, which breaks the chiral symmetry of QCD. The low energy excitations are pseudoscalar pions, with mass given by $m_\pi ^2 \simeq  m_q \Lambda_{\rm QCD}$ and baryons with mass $m_B \simeq \Lambda_{\rm QCD}$ (for a textbook review, see \cite{Schwartz:2013pla}). The effective field theory has the form (see e.g., \cite{Weinberg:2009bg}), 
\begin{equation}
\label{eq:QCD}
    \mathcal{L} = i \bar{N} \slashed{D}N + \frac{1}{2}\partial \Sigma\cdot\partial \Sigma - \frac{1}{2}m_{\pi}^2\,\Sigma\cdot\Sigma - m_B \bar{N} N + \mathcal{L}_{int},
\end{equation}
where $\Sigma$ a vector of pions and $N$ a baryon. The interaction Lagrangian is generated by residual strong force, and includes pion-pion, baryon-baryon, and baryon-pion interactions. The interactions of the pions are governed by the approximate chiral symmetry, under which the pions are the pseudo-Nambu-Goldstone bosons. The baryons and pions interact via derivative interactions, e.g.,
%\begin{equation}
%\label{eq:piN-int}
    $\mathcal{L}_{NN\pi} \simeq g_{\pi N} \frac{\partial_\mu \pi}{f_{\pi}} \, \bar{N} \gamma^\mu \gamma^5 N$,
%\end{equation}
where $g_{\pi N}$ is an interaction strength, and $f_\pi \simeq \Lambda_{\rm QCD}$ the pion decay constant. The pion mass is suppressed relative to the baryon mass by $m_{\pi} ^2 /m_B ^2 = m_q/\Lambda_{\rm QCD}$. In analogy with electrically millicharged dark matter \cite{HOLDOM198665,MOHAPATRA1990593}, the baryon-pion interaction allows dark baryons to emit pions by bremsstrahlung. However, we note the coupling $g_{\pi N}$ may be made arbitrarily small, unlike the analog in the Standard Model, where Weak decay of the neutron generates a large coupling of Standard Model nucleons to Standard Model pions via the Golderberger-Treiman relation \cite{coleman_1985,weinberg_1996}. We discuss this point further in App.~\ref{app:nucleonscattering} .

Moving to higher densities, while remaining at low temperatures,
%along the $x$-axis of Fig.~\ref{fig:QCD-phase-diagram}, 
and crossing into number densities larger than the mass of the lightest baryon, the vacuum of the theory changes \cite{Rajagopal:2000wf,Alford:2007xm}. For QCD-like theories with a low QCD scale, these phases can be cosmologically relevant, analogous to the wave-like nature of ultralight (``fuzzy'') bosonic dark matter \cite{Hu:2000ke,Hui:2016ltb}. Quantitatively, for $m_{\rm B} \lesssim 0.1 {\rm eV}$, the central regions of dark matter halos are at nuclear density. For example, in dwarf galaxies, where $\rho_{\rm DM} \sim 10^{8} M_{\odot}/{\rm kpc}^3$ \cite{Read:2018fxs}, or the inner region of the Milky Way, where $\rho_{\rm DM} \sim 1 - 10^3 {\rm GeV}/{\rm cm}^3$ ($\approx 10^7 - 10^{10} M_{\odot}/{\rm kpc}^3$) \cite{Lisanti:2016jxe}. One expects the dark matter in these regions is naturally in the quark condensate phase.

The effective action for quarks at non-zero chemical potential is given by Eq.~\eqref{eq:QCD} with an additional chemical potential term $\bar{q}\mu\gamma_0 q$. Single-gluon exchange generates an effective attractive 4-fermion interaction \cite{Alford:1998mk,Alford:1997zt, Alford:2007xm}, leading to a $\langle qq \rangle$ condensate,
\cite{Schafer:2002ms}
\begin{equation}
 \langle {q^{ i}_{\alpha a}q^{j}_{\beta c}\epsilon^{a c}} \rangle =  \Delta^{ij}_{\alpha\beta} = \Delta \epsilon^{ij} \epsilon_{\alpha \beta} = \Delta(\delta ^{i} _\alpha \delta^j _\beta - \delta^i _\beta \delta^j _\alpha ) , 
\end{equation}
where $i,j$ are flavor indices, $a,b$ are color indices, and $\alpha,\beta$ are spinor indices. The gap, $\Delta$, is the order parameter for the condensate phase. Physically, this condensate is the QCD analog to the Cooper pair of BCS theory \cite{PhysRev.106.162}. Quantitatively,  the gap is given by \cite{Son:1998uk,Alford:2007xm},
\begin{equation}
\label{eq:gap}
\Delta \simeq \frac{10^5 \mu}{g^5} e^{ - \frac{3 \pi^2}{\sqrt{2} g}},
\end{equation}
where $g$ is the gauge theory coupling. Note the gap, which sets the energy of a Cooper pair, can be much lower than the mass of a baryon. 

The $\langle q q \rangle$ condensate breaks the $SU(N)$ gauge symmetry, making this state a {\it superconductor}, along with the $U(1)_B$ baryon symmetry, making this state additionally a baryon {\it superfluid}. The flavor and color symmetries are broken to the diagonal subgroup, and this phase is thus termed the ``color flavor locking'' phase of QCD \cite{Alford:2007xm}. This simultaneously breaks chiral symmetry, leading to a spectrum of Goldstone bosons analogous the pions of the hadronic phase. Incidentally, recent work \cite{Reddy:2021rln} has considered a {\it lepton} superfluid; that model differs from STUMPs in that the fundamental degree of freedom is a scalar field \cite{Reddy:2021rln}, and not dark quarks, and the scalar has unit charge under lepton number, not a millicharge under baryon number.

This phase of SU(2) theory was proposed as dark matter candidate in \cite{Alexander:2018fjp}. A similar idea was proposed by \cite{Alexander:2016glq}, arising from a phenomenological four-fermion interaction. The physics is general to $SU(N_c)$ theories; however for this work, we specialize to a $N_c=3$ and $N_f=3$ species of light quark, analogous to the up, down, and strange quarks in color-flavor locking of visible QCD \cite{Alford:2007xm}. This allows us to utilize the decades of work in dense phases of the Standard Model QCD. In particular, the equation of state of dense quark matter is given by \cite{Alford:2004pf}
\begin{equation}
\label{eq:EOS}
    \Omega = - \frac{3}{4 \pi^2} a_4 \mu^4 + \frac{3}{4 \pi^2} a_2 \mu^2 + B_{\rm eff},
\end{equation}
where $\Omega$ is the free energy. The $a_4$ term corresponds to a gas of non-interacting quarks, while $a_2$ is generated by Cooper pairing and superconductivity, and $B_{\rm eff}$ is a free parameter that parametrizes all $\mu$-independent contributions to the free energy.

In our setup, as in visible QCD at extreme density, there is a Cooper pairing of quarks charged under the dark $SU(3)$, leading to color-flavor locking and an energy gap Eq.~\eqref{eq:gap}. Unlike visible QCD and neutron stars, there is no analog of the electroweak interactions, and thus all the physics are determined by the dark strong force, and hence, at high densities, the equation of state Eq.~\eqref{eq:EOS}. Following the philosophy of neutron star physics, we consider the parameters in the equation of state to be free parameters to be fixed by data, subject to the self-consistency condition that the EOS Eq.~\eqref{eq:EOS} only be applied in regions of space at nuclear density, i.e., regions wherein the effective field theory of the color-flavor locking phase is valid. 

Finally, the genesis of the dark baryons (STUMPs), implemented via an existing cogenesis scenario (e.g., decay of scalar carrying baryon number), may have produced the dark sector as a quark-gluon plasma, as hadrons, or as a quark condensate. If the initial genesis occurs to highly relativistic particles, the initial state can be expected to a thermal plasma. In this case, one might expect a relic background of dark pions; however, as suggested in \cite{Bai:2013xga}, the dark pions may decay into standard model particles. On the other hand, if the genesis produces a state with large chemical potential, and is below the diagonal of Fig.~\ref{fig:QCD-phase-diagram}, the initial state will be a condensate, that transitions to the hadronic phase as the universe expanse, and later re-enters the condensate phase in dense structures.  A third possibility is that the DM is  initially in the form of heavy hadrons, which subsequently decay to the light dark baryons. We expect these varied possibilities to be constrained by consistency with the cosmic microwave background, however, in this work we will focus on the late universe.

\section{The Tremaine-Gunn Bound}

The dark baryon (STUMP) proposed here is by nature a fermion. It is well known that there is a lower bound on the possible mass of a fermionic dark matter candidate \cite{Tremaine:1979we,Dalcanton:2000hn,Boyarsky:2008ju,Domcke:2014kla,DiPaolo:2017geq,Savchenko:2019qnn,Alvey:2020xsk}, which naively would rule out ultra-light dark baryons as dark matter. 

The simplest hint for such a bound is the inherent tension between the existence of heavy dense objects and the Pauli Exclusion Principle. More concretely, one may demand that the number density is large enough to constitute the total mass of a dwarf spheroidal galaxy, whilst simultaneously occupying only those states with velocity less than the escape velocity of the galaxy. This leads to a bound $m_{\rm DM} \gtrsim 100 \, {\rm eV}$; for example, the Fornax galaxy implies $m_{\rm DM} > 164 \, {\rm eV}$ \cite{Boyarsky:2008ju}.

The condensate phase of ultra-light QCD evades the escape velocity bound by construction: the Cooper pairing of quarks occurs between a quark with momentum $+\vec{k}$ and a quark with momentum $-\vec{k}$ to produce a zero-momentum Cooper pair. The resulting Cooper pair has, by definition, momentum far below the escape velocity of the galaxy, and is in no risk of escaping.

A less heuristic lower bound on the DM fermion mass is to note that the phase-space distribution function of dark matter, in the absence of collisions or dissipation, obeys conservation laws that follow from the Liouville's theorem. This approach is known as the Tremaine-Gunn bound \cite{Tremaine:1979we}. In particular, the maximum of the distribution function is conserved under time evolution. Assuming the fermions initially follow a Fermi-Dirac distribution, and assuming that the fermions eventually reach an isothermal distribution of radius $r_c$ and velocity dispersion $\sigma$, one finds the lower bound: $m_{\rm DM}^4
  \geq \frac{9(2\pi\hbar)^{3}}{(2\pi)^{5/2}g_s G_N\sigma \, r_{c}^{2}}.$
This bound is comparable to that following from the Pauli Exclusion Principle and escape velocity: observations of Fornax give $m_{\rm DM} \geq 195 \, {\rm eV}$ \cite{Boyarsky:2008ju}, and observations across a range of galaxies indicate $m_{\rm DM} \gtrsim 200 \, {\rm eV}$.

In the context of ultra-light QCD with dark matter comprised of dark baryons, the maximum of the distribution is not conserved under time evolution, due to the short-wavelength strong-interactions that mediate collisions, and bremsstrahlung of pions. Independent of the precise impact of these effects, the density of the final state is ultimately determined by nuclear physics, following the equation of state Eq.~\eqref{eq:EOS}. In what follows, we compute the pressure and density profile, calibrated by observations of dwarf galaxies.

\section{Condensate Cores in Dark Matter Halos}

The STUMP model proposed here sits at the intersection of ultralight dark matter \cite{Hu:2000ke,Hui:2016ltb,Ferreira:2020fam} and self-interacting dark matter (SIDM) \cite{Spergel:1999mh,Dave:2000ar}. Guided by those scenarios, one expects the formation of cores in the interior regions of galaxies, wherein the quark matter reaches a maximal density, in analogy to neutron stars. This addresses the core-cusp problem of collisionless cold dark matter, namely, observations favor cored halos with a maximum density, and not the cuspy halos predicted by collisionless cold dark matter, wherein the density is peaked at the center of the halo. (For a review and detailed discussion, see \cite{Bullock:2017xww}). We note that STUMPs are distinguished from past works on cored dark matter halos from light fermions  \cite{Randall:2016bqw,Domcke:2014kla} by the (strong) interactions.

\begin{figure}[h!]
    \centering
    \includegraphics[width=0.49\textwidth]{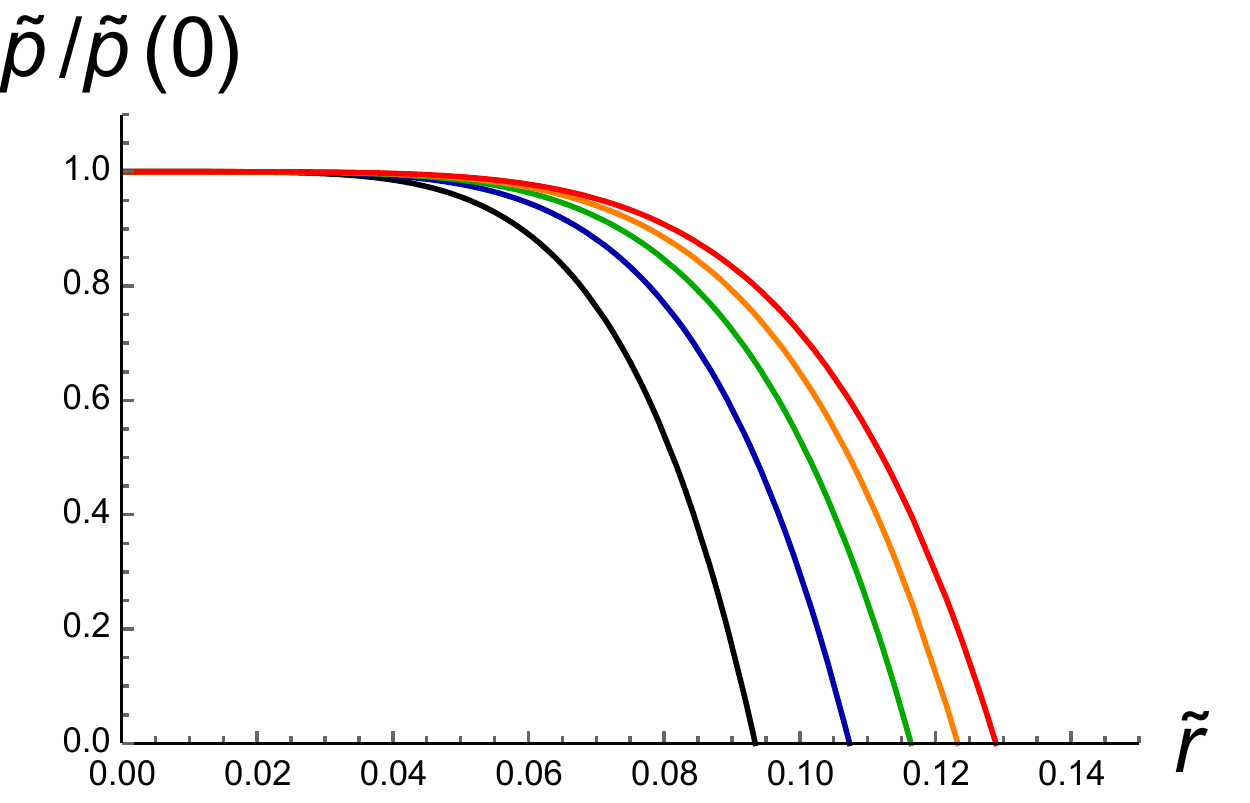}
    \caption{Family of dense dark QCD cores, expressed in dimensionless variables, and normalized to unity at the origin. Each solution has a finite radial extent, and terminates at $r=r_c$ defined by $p(r_c)=0$. Inside the core, i.e. for $r< r_c$, the density is very nearly constant in all cases. The differing solutions are distinguished by the radius of the core and hence total mass contained therein. Fixing the rescaling parameter $r_0 =1.5$ kpc, i.e., $r=1.5\tilde{r}$ kpc, we find the cores have radii $\sim 100 $ pc and total mass $\approx 10^8-10^9 M_{\odot}$. The green curve has properties in close agreement with the Tucana dwarf galaxy: $\rho_{\rm DM}(150 {\rm pc}) = 5.5 \times 10^{8} \, M_{\odot}/{\rm kpc}^3$ and core radius $r_c \simeq 176 {\rm pc}$ \cite{2019MNRAS.485.2010G}. }
    \label{fig:density-profiles}
    %\vspace{-1.0cm}
\end{figure}

In analogy with SIDM, the emission of light pions (which are parametrically lighter than the baryons) allows the halo to cool as it grows more dense. The result is a dense core of dark quark matter. The density and pressure of this core is determined by QCD physics that has been studied in the context of neutron stars. In contrast with SIDM, the cores are prevented from core-collapse by the pressure of the condensate phase, and thus we expect the cores to be long-lived.

Astrophysical observations of dwarf galaxies indicate a central density $\rho \sim 10^{8} M_{\odot}/{\rm kpc}^3 = 0.55 M_{\odot}/{\rm pc}^3 = (0.11 \, {\rm eV})^4 $ at a radius of $\sim 100$ {pc}.  Translating to natural units, these correspond to nuclear density ($n_{\rm DM}\sim m_{\rm DM}^3$) for $m_{\rm DM} \lesssim 0.1 \, {\rm eV}$. Note it is not necessarily the case that nuclear density triggers the onset of quark pairing; we expect a complicated structure to the outer edge of the condensate core, similar to visible neutron stars.  For simplicity in what follows we treat halos as an inner region described by the equation of state Eq.~\eqref{eq:EOS} and an outer region wherein the dark baryons behave as collisionless cold dark matter. The size and density of the condensate core are governed by the equation of state Eq.~\eqref{eq:EOS}, along with the Tolman-Oppenheimer-Volkoff (TOV) equations for hydrostatic equilibrium.

The TOV equations are given by,
\begin{eqnarray}
\frac{{\rm d}p}{{\rm d}r} &=& - \frac{G}{c^2} \frac{(p+\rho)(m+4\pi r^3 p/c^2)}{r^2[1-2Gm/(rc^2)]} \nonumber \\
\frac{{\rm d}m}{{\rm d}r} &=& 4\pi r^2 \frac{\rho}{c^2}.
\end{eqnarray}
The relevant thermodynamic relations are $p = - \Omega$, $n = \frac{{\rm d}\Omega}{{\rm d}\mu}$, and $\rho = \Omega + n \mu$,
from which we find the equation of state,
\begin{eqnarray}
 p[\rho] =&& \frac{1}{3} \left( \rho - 4B_{\rm eff}\right)  \nonumber \\+ &&\frac{r_{2,4}^2}{12 \pi^2}\left(-1 + \sqrt{1 + \frac{16 \pi^2}{r_{2,4}^2}(\rho - B_{\rm eff})} \right),
\end{eqnarray}
where we define $r_{2,4} \equiv \frac{a_2}{\sqrt{a_4}}$. We are interested in the regime wherein the physics are predominantly governed by the Cooper pairing of quarks, corresponding to $r_{2,4}\gg 1$.

To numerically solve this system we rescale all variables as $X = X_0 \tilde{X}$, with $\tilde{X}$ the dimensionless $X$. We define $
r= r_0 \tilde{r}$, $m=m_0 \tilde{m}$, $p=p_0 \tilde{p}$, $\rho= \rho_0 \tilde{\rho}$, where $r_0 = \frac{m_0}{24 \pi M_{pl}^2}$ and $p_0 = \rho_0 = (24 \pi)^3 \frac{M_{pl}^6}{m_0 ^2} $,  where we used $8 \pi G/c^4 = 1/(3 M_{pl}^2) $ with $M_{pl}=2.435 \times 10^{18} {\rm GeV}$, and we take $c=1$. With these substitutions, the TOV equations take a dimensionless form.

We find numerical solutions, which are shown, in dimensionless variables, in Fig.~\ref{fig:density-profiles}. These solutions have $r_{2,4} \gg 1$, though we note cores also exist in the opposite limit. We find that the central density is largely uncorrelated with the total mass, consistent with observations of dwarf galaxies. The radius is set by $r_c \propto \log (\rho_c/2 B_{\rm eff} - 1)$, and the total mass is $M \sim r_c ^3 \rho_c$. In all cases the central density is $\rho_c \simeq 2 B_{\rm eff}$, to a degree of fine-tuning determined by the total mass of infalling material.

Additionally, the frictionless transport that characterizes the high density (superfluid) environment, along with the confinement of the dark strong force into baryons that characterizes low density environments, allows the STUMP model to naturally satisfy constraints on dark matter self-interactions from merging galaxy clusters, e.g. from  the Bullet cluster \cite{Clowe:2003tk}.  More concretely, scattering of hadronic phase nucleons via pion exchange constrains the pion-nucleon coupling $g_{\pi N}$, which in turn is controlled by the lifetime of the dark `neutron'. The latter is stable (and hence the lifetime is infinite) in the limit of equal dark quark masses. We find that standard SIDM constraints (see \cite{Tulin:2017ara} for review) on the cross-section for $2\leftrightarrow 2$ nucleon scattering, $\sigma_{N N}$, impose $g_{\pi N}< 0.01$, which is easily satisfied in the STUMP model. This is discussed further in App.~\ref{app:nucleonscattering} .

 An interesting question is what happens when a dwarf galaxy falls into a halo, i.e, what happens when the condensate core hits hadronic phase matter (dark baryons) at velocities exceeding hundreds of kilometers per second. For a wide range of parameters, the baryon mass is greater than the value of the gap, and thus one expects the baryon (which, as a color singlet, is neutral relative to the superconductor) to act as a heavy impurity, namely, for the fluid to exhibit frictionless and dissipationless flow.  To the extent that the baryon can be treated as a superposition of three quarks, one might also expect scattering of quarks in the baryon with those in Cooper pairs \cite{Juzaki:2020ley,Sadzikowski:2002sk}, and scatter via a phenomenon known as Andreev Reflection \cite{Andreev1964THERMALCO}. As well studied in the context of superconductivity, an incident particle with energy greater than the energy gap $\Delta$ will pass through unimpeded, while a slow-moving charged particle will scatter via Andreev Reflection. For a wide range of parameter space, the velocity $200 {\rm km/s} \sim 10^{-3} {\rm eV}$ is above the energy scale of the condensate: e.g., for $g=0.1$, the gap is $\Delta = 10^{-81} \mu$ eV, while for $g=1$, the gap is $10^{-4} \mu$ eV, where $\mu \gtrsim m_{\rm DM}$, and thus one expects that for ultralight dark baryons only a small fraction of them to scatter with the superconducting core.

The small amount of scattering of dark baryons with the superconducting core will manifest on astrophysical scales as an additional drag force, or in analogy to dynamical friction, an `anomalous friction'. To estimate this effect, we take a fraction $f_{\rm scat}$ of the hadronic phase matter to be an optically thick medium through which the superconducting core is travelling. The total force acting on the core is,
\begin{equation}
    F \sim f_{\rm scat}\rho_{\rm halo} A v_{\rm halo}^2 ,
\end{equation}
where $f_{\rm scat}$ is the fraction of dark baryons undergoing scattering, and $A$ is the surface area of the core. From Newton's second law applied to the core, $F= M a \sim R_c^3 \rho_{\rm nucl} a$,  we deduce the time-scale of the anomalous drag,
\begin{equation}
t_{\rm drag} \equiv \frac{v}{a} \simeq \frac{\rho_{\rm nucl}}{f_{\rm scat}\rho_{\rm halo}} t_{\rm crossing},
\end{equation}
where $t_{\rm crossing} = R_c/v$ is the crossing time of the core. For $f_{\rm scat} \ll 1$, we expect $t_{\rm drag} \gtrsim t_{\rm crossing}$ and thus negligible anomalous friction on the core.

\section{Other Observational Signatures} 

Grounded in the physics of neutron stars, we have proposed here the STUMP scenario for dark matter, wherein dark matter is an ultralight fermion charged under standard model baryon number. This model starts from a loophole in the Tremaine-Gunn bound on fermionic dark matter, and ends with a prediction of near-constant density cores in dark matter halos. It fits into the paradigm of asymmetric dark matter as a millicharged dark baryon. 

This ultralight dark matter candidate is easily distinguished from its bosonic cousins. The gravitational and astrophysical signatures are distinguished from ultra-light axions by superconductivity, e.g., the existence of superconducting vortices \cite{Eto:2013hoa} and rotons, as well as Andreev reflection \cite{Andreev1964THERMALCO}, both of which arise in galaxy mergers. In particular, dynamical friction induces a transfer of angular momentum from the infalling galaxy to the host galaxy, suggesting the spontaneous production of superconducting vortices in the host. Other substructures, such as a `dark disk', may also form in this scenario, in analog to electrically millicharged dark matter \cite{Fan:2013tia,Fan:2013yva} and dark matter superfluids \cite{Alexander:2019qsh}, and may be amenable to detection via strong lensing \cite{Alexander:2019puy}.

The effective theory of interactions with the standard model is very different from an ultra-light boson. For example, we can expect effective 4-fermion interactions with the standard model fermions, of the form,  $\mathcal{L} = \frac{g_{Nf}}{\Lambda_B^2} \bar{N} N \bar{f} f$, where $N$ is the dark baryon, $f$ is a SM fermion, and $g_{Nf}$ is a coupling constant, allowing for 2-to-2 scattering. This can arise if baryon number is gauged with a mediator of mass $\Lambda_B$ (which satisfies $\Lambda_B >$TeV for consistency with collider experiments and proton decay). More generally, the effective theory of fermionic dark matter interactions is given in \cite{Fitzpatrick:2012ix}.

One might also expect an interaction between the dark pions and visible pions or baryons. Dark matter-baryon interactions have been studied in a number of cosmological contexts, e.g.~\cite{Munoz:2018pzp}. In contrast with the millicharged dark models considered there, the interaction in the millicharged dark QCD is suppressed by $\Lambda_B$, which must be $>$TeV for consistency with collider experiments and proton decay, and thus one does not expect any observable signal to be generated.

These interactions make predictions for both direct detection and collider searches for dark matter. We expect that direct detection searches for ultralight dark matter premised upon the interaction with electromagnetic fields
(e.g., resonant cavities) will return null results. On a more positive note, we expect a collider signature in the form of `semi-visible jets' \cite{Cohen:2015toa,Cohen:2017pzm,Cohen:2020afv}. The interaction with visible QCD will cause a fraction of visible QCD jets to be converted to dark QCD jets that subsequently pass through the detector undetected, rendering the visible QCD jets `semi-visible'. Further study will be needed to consider the ultralight dark baryon regime of this phenomenon.

 An additional intriguing possibility is early structure formation generated by the dissipation of energy and cooling of structures via the emission of dark pions, in analogy with self-interacting dark matter \cite{Pollack:2014rja,Essig:2018pzq}. Cooling by emission of light particles naturally leads to early structure formation, which may explain future observations of high redshift quasars at James Webb Space Telescope \cite{2019BAAS...51c.121F}. An analysis of cooling in general dissipative dark matter  models was performed in \cite{Essig:2018pzq}, which can be used to estimate the effect in the model proposed here. For STUMP dark matter this, excludes a small region of quark masses, $m_q/\Lambda_{\rm QCD} \simeq [10^{-18},10^{-14}]$. The details of this calculation are included in Appendix C. A more detailed analysis of parameter space, and minimal model extensions, may reveal interesting scenarios for STUMP dark matter. We leave this possibility to future work.

Turning to the early universe, the cogenesis of the STUMP and visible sector asymmetries may have its own observational signatures. For example, the production of fermions during inflation, associated with gravitational baryogenesis \cite{Alexander:2004us}, has a distinct signature in CMB polarization \cite{Lue:1998mq}. This mechanism has already been shown to produce the observed dark matter density in the closely related $SU(2)$ gauge theory with massless Weyl fermions \cite{Alexander:2018fjp}, and it will be interesting to generalize to $SU(3)$ with Dirac fermions, as proposed here. A final, tantalizing, possibility is that the vacuum energy of the dark condensate could act as dark energy \cite{Alexander:2016xbm,Addazi:2016nok}. 

\vspace{1cm}

\begin{acknowledgments}
{\bf Acknowledgements} The authors thank Gordan Z. Krnjaic, Jorge Sofo, Ken Van Tilburg, Liantao Wang, and Yiming Zhong, for useful discussions and insightful comments.
\end{acknowledgments}

%\newpage 
\appendix

\section{Schematic Phase Diagram of QCD}
\label{app:schematic}

\begin{figure}[h!]
%\vspace{1cm}
\begin{center}
\includegraphics[width=0.49\textwidth]{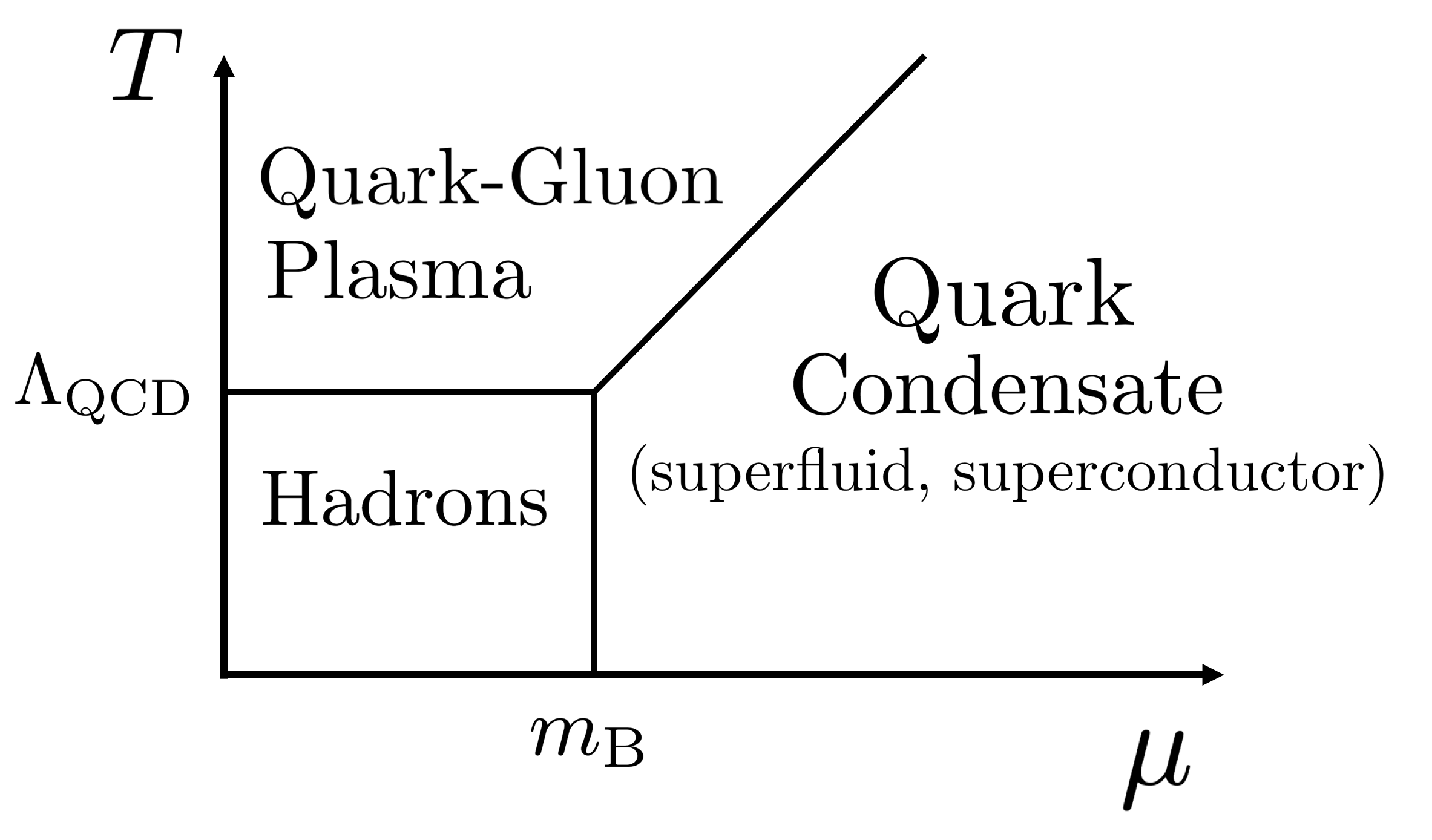}
\end{center}
\vspace{-.5cm}
\caption{A schematic phase diagram of QCD and QCD-like theories.  For a detailed phase diagram, see, e.g., \cite{Akiba:2015jwa}. The hadronic phase occurs at low temperature and low densities, while at densities above the mass of the lightest baryon $m_B$ a quark condensate forms, and at temperatures above the strong coupling scale $\Lambda_{\rm QCD}$, a quark-gluon plasma is formed.}
\label{fig:QCD-phase-diagram}
\end{figure}

\section{Milli-B-Charged Dark Matter}

In this appendix we build a concrete model of dark matter millicharged under standard model baryon number.  The starting point is to consider extensions of the standard model wherein  baryon number is a local symmetry, as has been done in many works, e.g.~ \cite{Rajpoot:1987yg,Foot:1989ts,Carone:1995pu,Dong:2010fw,FileviezPerez:2011pt,FileviezPerez:2011dg,Dulaney:2010dj,FileviezPerez:2010gw,Duerr:2013dza,Ma:2020quj}. The basic model requirements are (1) the cancellation of anomalies, and (2) consistency with observational constraints from proton decay.

In the Standard Model a local $U(1)_B$ symmetry is anomalous. In the standard notation of triangle diagrams, these are $SU(2)^2 U(1)_B$ and $U(1)_Y^2 U(1)_B$. These can be cancelled by adding an additional generation of fermions \cite{Duerr:2013dza}. Recently, building on \cite{Duerr:2013dza}, a class of anomaly-free gauged $U(1)_B$ models was given by \cite{Ma:2020quj}. This comprises new vector-like fermions $X^0$, $X^-$, and an SU(2) doublet $(E^0, E^-)$. The $U(1)_B$ charge assignments are determined by a free parameter $B_f$, given in Tab.~\ref{tab:B}, and all anomalies of the theory cancel for arbitrary values of $B_f$.

A simple model building framework is to introduce a complex scalar $S$ with integer baryon number.  In this case the coupling to standard model quarks is necessarily through irrelevant operators, which implies that at low energies baryon number is an approximate global symmetry. The violation of $U(1)_B$ occurs only through operators suppressed by powers of the $U(1)_B$ symmetry breaking scale, along with the conventional standard model global $U(1)_B$ anomaly. The model then easily satisfies constraints from proton decay. For a detailed discussion, see e.g.~\cite{Duerr:2013dza}.

\begin{table}[h!]
\centering
\begin{tabular}{|c|c|c|c|}
\hline
new fermion & $SU(2)_L$ & $U(1)_Y$ & $U(1)_B$ \\
\hline
$(E^0,E^-)_L$ & $2$ & $-1/2$ & $B_f$ \\ 
$(E^0,E^-)_R$ & $2$ & $-1/2$ & $B_f+3$ \\
\hline
$X^-_R$ & $1$ & $-1$ & $B_f$ \\
$X^-_L$ & $1$ & $-1$ & $B_f+3$ \\
\hline
$X^0_R$ & $1$ & $0$ & $B_f$ \\
$X^0_L$ & $1$ & $0$ & $B_f+3$ \\
\hline
\end{tabular}
\caption{New fermions transforming under $U(1)_B$ in the model of \cite{Ma:2020quj}. }
\label{tab:B}
\end{table}

All in all, an extension of the baryonic sector of the SM with gauged $U(1)_B$ is given by,
\begin{eqnarray}
    {\cal L} = && |D S|^2 + \frac{1}{4}B_{\mu \nu}B^{\mu \nu}  \\
    && - \lambda_s (S^2 - v_s^2)^2 + {\cal L}_{\rm new\, fermions}\nonumber 
    \end{eqnarray}
where here $S$ has baryon number $B_s \in \mathbb{Z}$. The  covariant derivative operator includes the $B$ gauge field, $B_{\mu}$. The $U(1)_B$ symmetry is broken when $S$ gets a VEV, but since $S$ has integer baryon charge, this is communicated to the standard model quarks only by higher-dimension operators \cite{Duerr:2013dza}, e.g., $\mathcal{L}\sim S^\dagger (eqqq)^{B_s}$. The Lagrangian ${\cal L}_{\rm new\, fermions}$ is the kinetic and mass terms for the new fermions (from e.g. \cite{Ma:2020quj}) required by anomaly cancellation.

It is straightforward to extend this to include a dark QCD. We emphasize that the SM B-anomaly is due to the $SU(2)$, and not the $SU(3)$. A dark SU(3) can be trivially extended to $SU(3)\times U(1)$, as follows
\begin{equation}
    {\cal L}_{DM} = |D \Phi|^2 + \frac{1}{4}F_{\mu \nu}F^{\mu \nu}+ i \bar{\chi} \slashed{D}\chi +\frac{1}{4} \lambda_\phi (\Phi^2 - v_\phi^2)^2
\end{equation}
where we introduce an additional complex scalar to generate a mass for the dark $U(1)$ at low energies.

Finally, the mixing of the standard model and dark matter is given by,
\begin{equation}
    {\cal L}_{mix} = \frac{\epsilon}{2} B_{\mu \nu} F^{\mu \nu}.
\end{equation}
Diagonalizing the kinetic mixing of the gauge bosons, one finds that DM has standard model baryon number $\epsilon/3$. This does not generate any anomalies, since the dark quarks are neutral under the SM electroweak interactions.

%%%%%%%%%%%%%%%%%%%%%%%%%%%%%%%%%%%%%%%%%%%%%%%%%%%%%%%%%%%%%%%%%%%

\section{Quark Condensation in QCD and QCD-like theories}

In this appendix we study in detail the pairing of quarks, and compute the finite-density induced ``gap'' $\Delta$ as a function of the gauge coupling $g$ and chemical potential $\mu$. We begin from the QCD Lagrangian at finite density,
\be \mathcal{L}  =  \bar{\psi}(i\gamma^{\mu} D_{\mu} - \mu \gamma^{0}) \psi - \frac{1}{4} {\rm Tr}\, G_{\mu\nu } {G^{\mu\nu}}  \ee
where $\mu$ is the chemical potential, which can be thought of as a proxy for the baryon number density, and ${G^{\mu\nu}}$ is the gluon field strength tensor, and the trace is over color indices.

The density-induced formation of a condensate may be computed rigorously in quantum field theory. To qualitatively understand this result, it is useful to first consider the effective field theory describing quarks at finite density. At finite density, the gluons obtain an in-medium mass, and may be `integrated out', leading to an effective 4-fermion interaction of the quarks \cite{Alford:1998mk,Alford:1997zt, Alford:2007xm}. More precisely, starting from the standard QCD interaction vertex 
\begin{equation}
    {\cal L}_{int}=\bar{\psi}_{i  \alpha} \gamma^{\mu } A_{\mu} ^a T^{a \alpha \beta}  \psi_{j   \beta} \delta^{ij} ,
\end{equation}
where $T$ is the $SU(N)$ generator, and using the $SU(N)$ identity,
\begin{equation}
     T^a _{\alpha \beta} T^a _{\gamma \delta} = \frac{1}{2}\left(\delta_{\alpha \delta} \delta_{\beta \gamma} - (1/N) \delta_{\alpha \beta} \delta_{\gamma \delta} \right),
\end{equation}
one finds the effective interaction for the quarks,
\be
 \label{eq:int} 
    \mathcal{L}_{int}  = g_{4f}\,  \bar{\psi}_{i \alpha } \gamma^\mu \psi_{j \beta} \bar{\psi}_{k \gamma} \gamma_\mu \psi_{l \delta} \delta^{ij} \delta^{kl}( 2 \delta^{\alpha \delta}  \delta^{\beta \gamma} - \delta^{\alpha \beta} \delta^{\gamma \delta}  ) 
 \ee
where $i,j,k,l$ are flavor indices, $\alpha, \beta, \gamma, \delta$ are color indices, and the spinor indices are suppressed. The coupling constant $g_{4f}$ is determined by the chemical potential and the gauge coupling. 

The interaction may be Eq.~\eqref{eq:int} may be attractive or repulsive depending on the color of the quarks. The interaction is attractive for $qq\rightarrow qq$ scattering when the incoming two-particle state has a  color wavefunction that is {\it antisymmetric} in color, i.e., the color wave function of the incoming state is $(|\alpha \beta\rangle - |\beta \alpha\rangle)/\sqrt{2}$. As in the BCS theory of superconductivity, the attractive interaction leads to the pairing of the fermions and the transition to the condensate phase of the theory. Furthermore, as in BCS, the pairing occurs for arbitrarily small values of the attractive coupling.

The formation of the condensate is encoded in the vacuum expectation value of a diquark state, namely, the gap $\Delta$. These are related by \cite{Schafer:2002ms}
\be
\label{eq:psiLpsiL} 
\langle {\psi^{ i}_{ \alpha a}\psi^{j}_{ \beta c}\epsilon^{a c}} \rangle =   \Delta \epsilon^{ij} \epsilon_{\alpha \beta} , 
\ee
where the $a,b$ indices are Dirac indices. The symmetry properties of the above, namely the index structure,  follows from the requirement that the state by antisymmetric in color, and that the wavefunction be totally antisymmetric.

The value of $\Delta$ may be computed at weak coupling by deriving and solving the `gap equation' \cite{Alford:2007xm}. Here we will perform this calculation at weak coupling.

The energy gap $\Delta$ is by definition a shift in the dispersion relation of excitations, $E_k^2 = \epsilon_k ^2+ \Delta_k ^2$. This manifests itself in field theory as the anomalous self energy of the fermion propagator, induced by interactions with gluons. The gap appears in the propagator as \cite{Kogut:2004su},
\be
\langle \psi^\dagger _a (p) \psi _a(p)\rangle = \frac{- i p_0 + \epsilon_p}{p_0^2 + \epsilon_p ^2 + \Delta^2_p}
\ee
where $\epsilon_p = |\vec{p}| - \mu$ and $a$ is the flavor index. This can be rearranged to compute $\Delta$ in terms of loop contributions to the propagator, leading to an expression for $\Delta$ that is an integral over quark-gluon interaction vertices and gluon propagators. For single gluon exchange, this takes the form
\be
\Delta \simeq g^2 \int  \mathrm{d}^4q \; v_\mu (q)  D_{\mu \nu}(q-k) v_\nu (-q)
\ee
where $v_\mu (q)$ is the (dressed) quark-gluon vertex and $D_{\mu \nu}$ is the gluon propagator. This is referred to as the ``gap equation.''

The gap equation for QCD may be rigorously computed from the two-particle irreducible (2PI) effective action.  The gap (or equivalently, the fermion self-energy) is given by the variation of the 2PI action with respect to the fermion propagator.  The gap equation for single gluon exchange is given by \cite{Alford:2007xm}
\be
\label{gapeq}
\Delta_{k} = \frac{g^2}{4} \int \frac{d^3 q}{(2 \pi^3)} \, Z(q) \frac{\Delta_{q}}{\epsilon_{q}} \left[ D_e(p) T_e + D_m (p)T_m \right] \ee
where $D_{e,m}$ are the electric and magnetic gluon propagators, and $Z$ is the wavefunction renormalization. The $T_{e,m}$ are numerical constants that come from traces over color, flavor, and Dirac indices.

The dominant contribution to the integral comes from gluons with $q_0 \ll q$, corresponding to the exchange of a gluon that is nearly static. Fixing the gauge to Coloumb gauge, the propagators in this limit are given by \cite{Alford:2007xm, Kogut:2004su}
\bea
D_{e}(q) &=& \frac{2}{|\vec{q}|^2 + m_{e}^2} \\
D_{m}(q) &=& \frac{1}{|\vec{q}|^2 +  (3 \pi/4)m_e^2 (q_0/|\vec{q}|)}
\eea
where $m_e \simeq g^2 \mu^2$ is the Debye mass. The Debye mass effectively screens electric gluons. The magnetic gluons, on the other hand, are damped rather then screened, on a characteristic scale $|\vec{q}| \sim (g^2 \mu^2 \Delta)^{1/3}$.

Focusing on the magnetic gluon contribution, the gap equation is of the form
\be
\Delta \simeq g^2 \int \mathrm{d}\xi \mathrm{d}\theta \; \frac{\Delta}{\sqrt{\xi^2 + \Delta^2}} \cdot \frac{\mu^2}{\theta \mu^2 + \delta^2} ,
\ee
where $\delta \simeq (g^2 \mu^2 \Delta)^{1/3}$ is the cutoff due to damping,  and $\xi\equiv k-\mu$, $\theta$ is the angle between the loop and external momenta. Performing the angular integration, the gap equation \eqref{gapeq} is given
\be
\Delta_k \simeq \frac{g^2}{18 \pi^2} \int \mathrm{d}q	 \; \frac{\Delta_q}{\epsilon_q} \frac{1}{2} \; \log \left( \frac{\mu^2}{|\epsilon_q ^2 - \epsilon_k ^2|}\right) .
\ee
The may be solved for $\Delta$ to give,
\be
\Delta \simeq g^{-5} \exp \left( - \frac{3 \pi^2}{\sqrt{2}g}\right) ,
\ee
This differs from the conventional BCS theory of superconductivity in the power of $g$ appearing in the exponent: the gap in BCS theory scales as $\exp(-1/g^2)$, and thus is highly suppressed relative to that in QCD and QCD-like theories.

\section{Nucleon-Nucleon Scattering}
\label{app:nucleonscattering}

The dark pion - dark nucleon interaction leads to nucleon-nucleon scattering in the hadronic phase. The resulting $2\leftrightarrow 2$ nucleon scattering is constrained by conventional SIDM constraints (see \cite{Tulin:2017ara} for a review).  In this appendix we quantify these constraints and discuss the implications for the model given theory expectations for the model parameters.

Making use of the Dirac equation for the nucleon, the pion-nucleon interaction may be recast as $ L_{\pi NN} \simeq  (m_N /f_\pi) g_{\pi N}\, \pi \bar N \gamma_5 N$.  Approximating $m_N \simeq f_\pi \simeq \Lambda_{\rm QCD}$ in the STUMP model, this can be further simplified to $L_{\pi NN} \sim g_{\pi N} \pi \bar N \gamma^5 N$. The resulting scattering cross section for nucleon-nucleon scattering is given by, 
\begin{equation}
    \sigma_{NN} \sim g_{\pi N} ^4 \frac{p^4}{s}\frac{1}{(s-m_\pi^2)^2}
\end{equation}
where $p$ is the
nucleon momentum, which can be approximated as  $p \simeq m_N v_N$, and $s$ is the center of mass energy squared.  In the limit $m_\pi \ll m_N$, this simplifies to
\begin{eqnarray}
    \frac{\sigma_{NN}}{m_N} && \sim g_{\pi N}^4 \frac{v_N^4}{m_N^3}  \\
    && \sim 1 \frac{{\rm cm}^2}{\rm g} \,  \left( \frac{g_{\pi N}}{0.01} \right)^4  \left(\frac{v_N}{10{\rm km/s}}\right)^4  \left(\frac{m_N}{0.1{\rm eV}}\right)^{-3} .
\end{eqnarray}
From the benchmark SIDM constraint $\sigma < 1 {\rm cm }^2/{\rm g}$ we find the constraint $g_{\pi N} < 0.01$ in the STUMP model. To understand the implications of this for the model, we must understand the dark QCD theory prediction for $g_{\pi N}$.

The coupling $g_{\pi N}$ is related to the decay of the ``neutron''. In the Standard Model, this is encapsulated by the Goldberger-Treiman relation \cite{coleman_1985,weinberg_1996}, which relates the pion-nucleon coupling to the lepton-nucleon coupling. The coupling of nucleons to leptons is parametrized by $g_V$ and $g_A$, for vector and axial nucleon-lepton interactions. In historical terms, the vector interaction with strength $g_V$ generates Fermi Decay of the neutron while the axial interaction with strength $g_A$ generates Gamow-Teller decay.  These can be further expressed in terms of the Fermi coupling and the CKM matrix elements, as $g_V = G_F V_{ud}$, and $g_A \equiv G_F V_{ud} \lambda$ with $\lambda \equiv g_A /g_V$.  

The Goldberger-Treiman relation is given by \cite{coleman_1985},
\begin{equation}
    g_{n \pi} = g_{A} \frac{m_N}{f_{\pi}}
\end{equation}
where $g_A$ is the axial coupling of nucleons to leptons. From this one may compute the neutron lifetime as,
\begin{equation}
    \frac{1}{\tau_N} \simeq |\langle p  | n \rangle|^2 \propto (g_V^2 +3  g_A^2)
\end{equation}
The limit of a stable neutron corresponds to vanishing $g_V$ and $g_A$.

In the Standard Model, measurements of neutron lifetime set $g_{\pi N} = \mathcal{O}(1)$. In the STUMP model, where there is no analog of the Weak force or of beta decay, the stability of the neutron is controlled only by the mass splitting of the dark quarks. For nearly-degenerate quark masses, we expect $g_{\pi N} \ll 1$, and thus the model easily satisfies  SIDM constraints. 

\section{Cooling Rate}

An analysis of cooling in general dissipative dark matter  models was performed in \cite{Essig:2018pzq}, which can be used to estimate the effect in the model proposed here.  The cooling rate is in general given by,
\begin{equation}
 \label{eq:bulk-cooling-rate}
     C  = \rho^2 \frac{\sigma_{\rm diss.}}{m_{\rm DM}} \frac{4\nu \nu_{\rm loss}^2}{\sqrt{\pi}} \left( 1 +\frac{\nu_{\rm loss}^2}{\nu^2} \right) e^{-\frac{\nu_{\rm loss}^2}{\nu^2}},
\end{equation}
where $\nu_{\rm loss}\equiv\sqrt{E_{\rm loss}/m}$ is the velocity loss per event, $\sigma_{\rm diss.}$ is the dissipative interaction cross section, and $\nu$ is the dark matter velocity dispersion, which for the Milky Way is roughly $\nu_{\rm MW} \sim 220\, {\rm km/s} \sim 10^{-3} c$ \cite{Bozorgnia:2016ogo}. Constraints on the cooling rate can be derived by assuming a Boltzmann distribution for $\nu$, and comparing the total cooling to the dynamical time scale of the halo  $t_{\rm dy}= H/\nu$ \cite{Essig:2018pzq}. The observational constraints depend on the relative size of the scattering cross section and the dissipation cross section, and are shown in Fig.~3 of \cite{Essig:2018pzq}.

In our setup, both scattering and dissipation arise solely from the residual strong force. Hence, the scattering and dissipation cross sections are roughly equal: $\sigma_{\rm diss.} \simeq \sigma_{\rm scat.}$. The energy loss per bremsstrahlung is the rest energy of a pion, $E_{\rm loss} \simeq m_{\pi} c^2$, from which we find the relative velocity loss $\nu_{\rm loss}$,
\begin{equation}
    \nu_{\rm loss} \equiv \sqrt{\frac{E_{\rm loss}}{m_{\rm DM}}} \simeq \sqrt{\frac{m_{\pi}}{m_{\rm DM}}} c.
\end{equation}
In our model, we can express the above in terms of the quark mass and strong coupling scale:
\begin{equation}
    \nu_{\rm loss} = \left( \frac{m_q}{\Lambda_{\rm QCD}} \right)^{\frac{1}{4}} c .
\end{equation}
From this we see that the cooling rate Eq.~\eqref{eq:bulk-cooling-rate} is suppressed in both the limit of massless quarks and for quarks heavier than $\approx 10^{-13} \Lambda_{\rm QCD}$,  the latter due to the exponential suppression in Eq.~\eqref{eq:bulk-cooling-rate}. For eV-scale dark baryons, this translates to $m_q \gtrsim 10^{-13}$ eV in order for cooling to be exponentially suppressed.

Indeed, comparing to the observational constraints, Fig.~3 of \cite{Huo:2019yhk}, for  $\sigma_{\rm diss.} \simeq \sigma_{\rm scat.}$ the data excludes $ \nu_{\rm loss}$ above $\sim 10$ km/s but below $\sim 100$ km/s, which for STUMP dark matter excludes a small region of quark masses, $m_q/\Lambda_{\rm QCD} \simeq [10^{-18},10^{-14}]$.

\section{Early Universe Physics of STUMP Dark Matter}
\label{app:early}

In this work we are particularly interested in the late universe physics of STUMP dark matter. It is important, however, to demonstrate that the model has a self-consistent early universe description, and is compatible with known constraints on dark matter and dark radiation, such as the CMB and big bang nucleosynthesis. Here we outline one simple early universe genesis mechanism of STUMP dark matter.

We first note early universe genesis of  an ultralight dark QCD condensate developed in \cite{Alexander:2018fjp}. In that work, the dark matter is in the condensate phase at very early times, and the relic density of dark matter is set by the energy density initially stored in the bosonic excitations of the condensate. The resulting cosmology is near-indistinguishable from a conventional axion. 

The connection to the present paper arises if the condensate, behaving as a coherent scalar field in the early universe, is fragmented in the late universe, causing the theory to enter the hadronic phase.  The fragmentation of scalar field dark matter had been studied recently in \cite{Cotner:2019ykd} . It will be interesting to study this in detail for the STUMP model, but we would opt to leave that for future work.

%%%%%%%%%%%%%%%%%%%%%%%%%%%%%%%%%%%%%%%%%%%%%%%%%%%%%%%%%%%%%%%%%%%

\bibliography{DM-NS-refs}
\bibliographystyle{JHEP}

\end{document}